\documentclass[
reprint,
superscriptaddress,
amsmath,amssymb,
aps,
prab,
floatfix,
]{revtex4-2}

\usepackage{graphicx}%
\usepackage{dcolumn}%
\usepackage{bm}%
\usepackage[english]{babel}
\usepackage[utf8]{inputenc}
\usepackage{siunitx}
\usepackage[dvipsnames]{xcolor}
\usepackage{xspace}
\usepackage{booktabs}
\usepackage{multirow}
\usepackage[normalem]{ulem}
\usepackage{hyperref}%
\hypersetup{
    colorlinks=true,
    linkcolor=blue,
    filecolor=magenta,      
    urlcolor=cyan,
}
\makeatletter
\renewcommand{\fnum@figure}{\normalfont FIG. \thefigure}
\renewcommand*{\@caption@fignum@sep}{\normalfont . }
\makeatother

\newcommand{\Ljpp}{L_{\mathrm{sp}}}
\newcommand{\Lpps}{L_{\mathrm{drift}}}
\newcommand{\rms}{\mathrm{rms}}

\newcommand{\rmscalc}{\rms,\,\mathrm{calc}}
\newcommand{\rmsin}{\rms,\,\mathrm{in}}
\newcommand{\rmsexp}{\rms,\,\mathrm{exp}}
\DeclareSIUnit\mrad{mrad}
\DeclareSIUnit\urad{\si{\micro\radian}}
\DeclareSIUnit\pixel{pixel}
\DeclareSIUnit\Bpercent{\mathrm{0.1}\si{\percent}}

\begin{document}

\preprint{APS/123-QED}

\title{Limitations of emittance and source size measurement of laser-accelerated electron beams using the pepper-pot mask method}%

\author{F.~C.~Salgado}
\email{felipe.salgado@uni-jena.de}
\author{A.~Kozan}
\author{D.~Seipt}
\author{D.~Hollatz}
\author{P.~Hilz}
\author{M.~C.~Kaluza}
\author{A.~S{\"a}vert}
\author{A.~Seidel}
\author{D.~Ullmann}
\author{Y.~Zhao}
\author{M.~Zepf}
\affiliation{Institute of Optics and Quantum Electronics, Friedrich-Schiller-Universit{\"a}t, Max-Wien-Platz 1, 07743 Jena, Germany}
\affiliation{Helmholtz-Institut Jena, Fr{\"o}belstieg 3, 07743 Jena, Germany}
\affiliation{GSI Helmholtzzentrum f{\"u}r Schwerionenforschung, Planckstraße 1, 64291 Darmstadt, Germany}
\date{\today}%

\begin{abstract}
The pepper-pot method is a widely used technique, originally proposed for measuring the emittance of space-charge-dominated electron beams from radio-frequency photoinjectors. With recent advances in producing high-brightness electron beams via laser wakefield acceleration (LWFA), the method has also been applied to evaluate emittance in this new regime. Here, the limitations of this method for measuring the emittance of LWFA electron beams are investigated, particularly in parameter regimes where the true beam emittance is overestimated.
Conducting an experiment at the JETi200 laser system, we measured an upper bound for the geometric beam emittance of $(26.2\pm7.3)$~\si{\um\mrad} using the pepper-pot method. This result is consistent with GEANT4 Monte Carlo simulation of the pepper-pot diagnostic, with an input beam-emittance parameter that matches both PIC simulations of the laser-plasma accelerator and an independent measurement using the transient optical grating method.
\end{abstract}

\maketitle

\section{Introduction}

Small source sizes and extraordinarily high-quality electron beams are fundamental requirements for driving the new generation of advanced light sources, including free-electron lasers (FELs)~\cite{Wang.2021, Fuchs.2009, Huang.2012, Maier.2012} and Thomson sources~\cite{Geddes.2015, Powers.2014} as well as electron-positron particle colliders~\cite{Leemans.2009, Schroeder.2010, ALEGROcollaboration.2019, Assmann.2020}. These applications demand normalized transverse emittances at the \si{\um\mrad}~level, for example, the FEL facilities requiring slice emittances as low as 100-200~\si{\um\mrad} to achieve optimal performance~\cite{Prat.2019}. An increasingly attractive alternative to radio-frequency (RF) accelerators for producing such high-quality beams is particle wakefield accelerators, including laser-driven (LWFA)~\cite{Leemans.2006, Esarey.2009, Gonsalves.2019, Picksley.2024} and particle-driven (PWFA) schemes~\cite{Chen.1985, Gotzfried.2020}. A laser wakefield accelerator generates longitudinal electric fields with gradients exceeding 100~GV/m, approximately three orders of magnitude higher than RF accelerators. This exceptional capability enables the generation of GeV beams with normalized transverse emittances below 0.5~$\pi~\si{\mm\mrad}$~\cite{Weingartner.2012, Kneip.2012, Golovin.2016} on a centimeter scale, whereas RF accelerators require close to 100 meters~\cite{Malka.2002}.

The quality of electron beams can be characterized by the brightness parameter $\mathcal{B}_n$, as described in Ref.~\cite{Reiser.2008}:
\begin{equation}
\mathcal{B}_n \propto \frac{I}{\epsilon_{n_x} \, \epsilon_{n_y}} \,,
\label{eq. brillance definition}
\end{equation}
which it is a function of the beam current $I$ and the transverse normalized emittances $\epsilon_{n_x}$ and $\epsilon_{n_y}$ in the $x$ and $y$ directions, respectively.

To achieve a high-brightness beam, one can either increase the beam current $I$ or decrease the emittance of the particle beam. In this study, we specifically focus on the transverse emittance parameter, a critical property of the beam describing how well the particles are confined within its transverse phase space. High-brightness beams can be achieved by methods such as laser-wakefield acceleration or plasma-photocathodes which exhibit brightness above $10^{17}~\si{\ampere/\meter^2/\Bpercent}$~\cite{Manahan.2017, Xiang.2024}. Therefore, accurate methods to measure the emittance of a beam are required to correctly determine the brightness of a particle beam.

While scanning methods such as quadrupole and solenoid scans are widely used to characterize the transverse emittance of electron beams on the order of a few~\si{\mm \mrad}, it is important to note that these methods cannot provide shot-to-shot information~\cite{Weingartner.2012, Prat.2014, Ji.2019}. They require multiple shots to construct a complete picture of the beam's emittance, which can be problematic for systems with significant shot-to-shot variations.

In contrast, single-shot techniques such as transverse deflecting structures (TDS)~\cite{Akre.2001, Behrens.2014, Dolgashev.2014, Floettmann.2014, Maxson.2017, Tan.2019, Marchetti.2021}, Shintake monitors~\cite{Tenenbaum.1999, Yan.2012}, and laser gratings~\cite{Seidel.2021, Salgado.2024} can capture emittance information for each individual electron bunch, making them more suitable for characterizing beams with high variability or for applications requiring real-time feedback. However, these methods introduce additional complexity to the emittance diagnostics, such as requiring extra magnets and optics in the setup.

To enable real-time measurements while minimizing diagnostic complexity, the pepper-pot (PP) method has been used for evaluating the emittance of laser wakefield acceleration (LWFA) beams. Its simplicity lies in requiring only a single pepper-pot mask, typically a few square millimeters in size, placed in the beam path of the experiment, with a scintillating screen positioned a few tens of centimeters downstream of the PP mask~\cite{Fritzler.2004, Shanks.2009, Sears.2010, Brunetti.2010, Manahan.2014}.

Initially, the pepper-pot method was primarily designed and extensively utilized to characterize the phase space of RF photoinjectors. These photoinjectors typically feature electron beams with a few-MeV energy and a large beam waist, resulting in low divergence~\cite{Rosenzweig.1994, Ganter.2010, Anderson.2002, Apsimon.2019}. Such beams are often dominated by space charge effects.

The propagation of an electron beam can be characterized as either space charge-dominated or emittance-dominated. To classify the regime, we can compare the terms of the root mean square (RMS) beam envelope equation in a drift space~\cite{Anderson.2002, Reiser.2008, Apsimon.2019},
\begin{equation}
    \sigma''_{x} = \frac{\epsilon_n^2}{\gamma^2 \sigma_x} + \frac{I}{\gamma^3 I_0 \left( \sigma_x + \sigma_y \right)} \,,
    \label{eq. space-charge or emittance dominated beam}
\end{equation}
where $I$ is the peak beam current, $I_0=17$~kA is the Alfv{\'e}n current, $\epsilon_n$ is the normalized emittance of the beam, and $\sigma_x$ and $\sigma_y$ are the RMS beam waist, also known as the source sizes, in the $x$- and $y$-directions. 

By taking the ratio $\mathcal{R}$ of the two terms on the right-hand side of Eq.~\eqref{eq. space-charge or emittance dominated beam}, and assuming a round beam, \textit{i.e.}, $\sigma_0 = \sigma_x = \sigma_y$, we obtain~\cite{Anderson.2002},
\begin{equation}
    \mathcal{R} = \frac{I\, \sigma_0}{2\, I_0\, \gamma\, \epsilon_n^2}\,,
    \label{eq. ratio space emittance dominated}
\end{equation}
which can be used to determine whether the electron beam is space-charge or emittance dominated. 

A beam is said to be space-charge dominated when its ratio $\mathcal{R} \gg 1$. Beams generated by RF-photoinjectors, for example, easily fall into this category due to their divergence of a few-\si{\mrad}, energy of about 5~\si{\MeV}, and typical RMS waist sizes of approximately 0.5~\si{\mm}~\cite{Abrahamyan.2003}. In contrast, LWFA electron beams~\cite{Couperus.2017, Li.2017} can achieve higher energies in the order of 100’s of MeV and beyond at the output of the first stage, with much smaller beam waists on the order of a few~\si{\um} while maintaining their divergence similar to radio-frequency accelerated beams of few-\si{\mrad}. This combination of parameters leads LWFA beams to have an $\mathcal{R} \approx 0.07 \ll 1$, indicating that they are emittance dominated.

In this work, we investigate the properties of the pepper-pot method for diagnosing the emittance of laser-accelerated electron beams, focusing on how accurately it can infer the beam's transverse emittance and waist. We also investigate the range of applicability of the pepper-pot technique in inferring the emittance of LWFA electron beams.

The paper is organized as follows. We briefly introduce the theory of beam emittance in Section~\ref{section:theory emittance} and the pepper-pot mask technique in Section~\ref{section:theory pp}. Then, we benchmark the pepper-pot method using Monte Carlo simulations in Section~\ref{sec. testing pp method} and show the limitations of the method in Section~\ref{section:limitation of pp method}. Then, we present experimental results in Section~\ref{section:experiment} that support our findings. Finally, we end with conclusions in Section~\ref{section:conclusions}.

\section{Theory}
In this section, we briefly introduce the definition of beam emittance and later the theory that underlies the method used to estimate the RMS emittance using pepper-pot masks.

\subsection{Beam emittance}
\label{section:theory emittance}
The emittance of a beam is defined as its volume occupied in phase space. For a beam without accelerating forces acting upon it, the emittance remains constant during its propagation (assuming ideal magnets and quadrupoles and neglecting radiation reaction). This conservation follows from Liouville's theorem, which states that phase space density is preserved in Hamiltonian systems with conservative forces. Ideal magnetic elements satisfy this condition because magnetic forces are always perpendicular to the particle's direction of motion. Since perpendicular forces perform no work on the particles, they can only change the particle's direction never its speed or energy, clearly making the system conservative. Neglecting radiation reaction is crucial because radiation emission would introduce energy loss and non-conservative forces, violating the conditions required for emittance conservation.

Let us consider the electron beam propagates along the $z$-axis, and restrict the discussion to one transverse spatial dimension.
The transverse phase space ellipse of a particle beam can be expressed the symmetric beam matrix
\begin{gather}
    \bm{\sigma} = \begin{pmatrix}
                        \langle x^2 \rangle & \langle x x^{\prime} \rangle\\
                        \langle x x^{\prime} \rangle & \langle x^{\prime\,2} \rangle
                  \end{pmatrix} \,,
\end{gather}
such that $\langle \bm{u}^{\mathrm{T}} \bm{\sigma}^{-1} \bm{u} \rangle = 1$, with the particle's coordinate vector $\bm{u} = (x, x^\prime)$, where $x$ represents the particle's position, $x^\prime$ denotes its divergence, and the notation $\langle \, \rangle$ represents the average of the quantity within the brackets. 

The phase space area is bounded by the density distribution of the particles $\varrho(x, \, x^\prime) = \, \varrho(\bm{u}^{\mathrm{T}} \, \bm{\sigma}^{-1} \, \bm{u})$,
\begin{equation}
    \varrho(x, \, x^\prime) \propto  \exp \left[ -\frac{\langle x^{\prime\,2} \rangle\, x^2 - 2\langle x x^{\prime} \rangle\,x\,x^\prime + \langle x^2 \rangle\,x^{\prime^2}}{2 \, \det \bm{\sigma}} \right] \,,
    \label{eq. particle distribution}
\end{equation}
assuming a Gaussian distribution. The 1/$\pi$ area occupied by the distribution above in Eq.~\eqref{eq. particle distribution}, \textit{i.e.}, the RMS emittance of the electron beam, is then defined as
\begin{gather}
    \epsilon_{\rms}^2 = \det \bm{\sigma} = \, \langle x^2 \rangle \langle x^{\prime\,2} \rangle - \langle x x^{\prime} \rangle^2 \, \approx\, \sigma_{x}^2 \, \theta_{x}^{2} \,,
    \label{eq. rms emittance definition}
\end{gather}
defining the RMS source size $\sigma_x =\sqrt{\langle x^2\rangle}$ and divergence $\theta_x =\sqrt{\langle x'^2\rangle}$. The subscript $x$ denotes the transverse axis along which the beam waist and divergence are evaluated.
The approximation in Eq.~\eqref{eq. rms emittance definition} of neglecting the correlation term $\langle x x^{\prime} \rangle^2$ is applicable to the laser–plasma accelerator used in this work, as its value is more than two orders of magnitude smaller than the leading term $\langle x^2 \rangle \langle x^{\prime\,2} \rangle$ at the plasma–vacuum interface, as shown by our particle-in-cell (PIC) simulations (see Appendix~\ref{appendix:pic} for details) under the experimental conditions of the plasma-accelerator stage as shown later in Section~\ref{section:experiment}.

An ideal beam, often called quasi-laminar, consists of particles traveling at very small transverse angles relative to the beam axis, with trajectories that do not intersect. Such a beam occupies minimal area in phase space and thus exhibits very low emittance.

On the other hand, a broadband particle beam is composed of particles with different energies and, consequently, the RMS emittance given in Eq.~\eqref{eq. rms emittance definition} does not remain constant throughout the propagation of the beam due to the chromatic emittance growth~\cite{Migliorati.2013, Seipt.2021}. To account for such effect, one can normalize the RMS emittance given in Eq.~\eqref{eq. rms emittance definition} with respect to the energy of the particles within the beam, here represented by the average Lorentz-factor $\langle \gamma \rangle$ of the particles. The normalized emittance $\epsilon_n$ is then evaluated as~\cite{Antici.2012}
\begin{equation}
    \epsilon^2_{n_x} = \, \langle \gamma \rangle^2 \left[ \left( \frac{\sigma_E}{E} \right)^2 \, \langle x^2 \rangle \, \langle x^{\prime\,2} \rangle + \epsilon_{\rms}^2\right]  ,
    \label{eq. norm emittance with energy spread}
\end{equation}
where the term $\sigma_E/E$ is the energy spread of the beam given as
\begin{equation}
    \left( \frac{\sigma_E}{E} \right)^2 = \, \frac{ \langle \beta^2 \, \gamma^2 \rangle - \langle \beta \, \gamma \rangle^2 }{\langle \gamma \rangle^2} \,.
    \label{eq: norm emittance}
\end{equation}

As the beam envelope is assumed to be at its waist at the plasma–vacuum interface, its transverse size increases due to free expansion during subsequent drift~\cite{Antici.2012}. Hence, for a sufficiently long propagation length in vacuum after the beam exits the plasma, one can express Eq.~\eqref{eq. norm emittance with energy spread} as~\cite{Antici.2012, Migliorati.2013}
\begin{equation}
    \epsilon^2_{n_x} = \, \langle \gamma \rangle^2 \left[ \left( \frac{\sigma_E}{E} \right)^2 \, \theta_{x}^4 \,  L^2+ \epsilon_{\rms}^2\right] \,,
    \label{eq. norm-emittance with drift}
\end{equation}
where the dependence on the drift length $L$ is now explicitly included such that the average RMS beam size is given by $\langle x^2 \rangle = \theta_{x}^2 \,  L^2$, and the average RMS beam divergence as $\langle x^{\prime\,2} \rangle = \theta_{x}^2$.

\subsection{The pepper-pot mask technique}
\label{section:theory pp}
The pepper-pot method was initially developed to diagnose space-charge-dominated beams by transforming the beam into beamlets that possess sufficient charge but are not significantly defocused due to space charge effects, \textit{i.e.}, emittance-dominated beamlets. The pepper-pot consists of a high-density material with a grid of holes or slits oriented perpendicular to the propagation axis of the electron beam. This mask samples the test electron beam into smaller, emittance-dominated beamlets, which then propagate toward a scintillation screen positioned downstream. The screen is then imaged using a high-resolution imaging system. Finally, the emittance of the electron beam is determined by analyzing the sizes and positions of the beamlets, as well as the hole diameter, pitch, and alignment of the mask in the experimental setup, following the mathematical framework introduced in Ref.~\cite{Zhang.1996}.

In this section, we briefly introduce the method and algorithm that we use to reconstruct and retrieve the RMS emittance of the test electron beams utilizing pepper-pot masks.

\subsubsection{Pepper-pot mask design constrains}
The defining properties of a pepper-pot mask are the diameter $d$ of the holes, the distance pitch $\Lambda$ between their centers, the mask thickness $\Delta z$ and its high-Z material. The material of the mask should be selected in a way that the particles propagating through it are absorbed or scattered at large angles, for example, Tungsten, which has a high stopping power. The thickness $\Delta z$ of the mask must be chosen such that it is larger than the radiation length of the material used, $\Delta z \geq X_0$. The radiation length $X_0$ is approximated as~\cite{Anderson.2002, Apsimon.2019},
\begin{gather}
    X_0 = \frac{E}{\frac{\mathrm{d}\,E}{\mathrm{d}\,z}} \approx  \frac{E \,(\si{\MeV})}{1.5 \, (\si{\MeV\cm\squared\per\gram}) \, \rho \,(\si{\gram}/\si{\cubic\cm})} \,,
\end{gather}
where $E$ is the energy of the incident beam, $\rho = 19.3~\si{\gram}/\si{\cubic\cm}$ is the density of Tungsten.

The next design rule regards the distance in which the screen should be placed downstream from the mask, $\Lpps$. To avoid overlap between the beamlets on the screen, the distance $\Lpps$ should be chosen such that the condition $4\theta_{x}\Lpps < \Lambda$ is fulfilled~\cite{Anderson.2002}.

Finally, the imaging system of the electron beam diagnostic should be capable of resolving the angular spread of the beamlets on the screen. Therefore, the position and angular resolution of the imaging system should be comparable, $\sigma_{x}/\Lambda = \Lpps \theta_{x}/ r_i$, where $\sigma_{x}$ stands for the size of the beam and $r_i$ is the resolution of the imaging system given in $\si{\um}/\si{\pixel}$.

These conditions are easily fulfilled for RF-photoinjector beams, which typically have beam waists around $0.5~\si{\mm}$ and an RMS divergence of the same order of magnitude, approximately $0.5~\si{\mrad}$. For instance, such beams have been adequately sampled by pepper-pot masks with $d=100~\si{\um}$, $\Lambda=200~\si{\um}$, and a screen placed approximately $60~\si{\cm}$ from the mask~\cite{Apsimon.2019}. 

In contrast, LWFA beams hardly satisfy all these design criteria simultaneously. Consequently, the precision of the emittance measurement of LWFA electron beams is an open question. In the following sections, we investigate the pepper-pot method with respect to LWFA parameters and show the limitations of the method to resolve the emittance of this class of electron beams.

\subsubsection{Evaluating the beam emittance}
\label{subsec:evaluating beam emittance}
After the interaction of beamlets with the screen, the emitted scintillation light is imaged, and the signal is integrated over the x- and y-axis. The integrated signal is then used to evaluate the RMS emittance of the electron beam by applying the following equation which uses the second central moment of the particle distribution $\langle x^2 \rangle$, $\langle x^{\prime\,2} \rangle$, and $\langle x x^{\prime} \rangle$ previously introduced in Eq.~\eqref{eq. rms emittance definition}~\cite{Zhang.1996, Sears.2010},
\begin{multline}
    \epsilon^2_{\rms} = \frac{1}{N^2} \left\{ \left[ \sum^p_{j=1} n_j (x_j - \overline{x})^2  \right] \right. \\
    \times \left[ \sum^p_{j=1} \left[ n_j \sigma_{x^{\prime}_j}^{\prime\,2} + n_j ( x^{\prime}_j - \overline{x}^{\prime} )^2  \right] \right] \\
    \left. -  \left[ \sum^p_{j=1} n_j x_j x^{\prime}_j - N \overline{x} \, \overline{x}^{\prime}  \right]^2   \right\} \, ,
    \label{eq. PP equation}
\end{multline}
where $N = \sum^p_{j=1} n_j$, with $n_j$ is the number of electrons propagating within the $j^{\mathrm{th}}$~beamlet, and $p$ as the total number of beamlets. 
Furthermore, $x_j$ is the centroid position of the $j^{\mathrm{th}}$~beamlet (in the slit plane, i.e. the location of the $j$-th hole), and $\overline{x} = \sum^p_{j=1} n_j x_j / N$ is the mean position of all beamlets. The RMS divergence of the $j^{\mathrm{th}}$~beamlet is given by $\sigma_{x^{\prime}_j}^{\prime}$, $x^{\prime}_j = \left( X_j - x_j \right)/\Lpps$ is the mean divergence of the $j^{\mathrm{th}}$~beamlet, where $X_j$ is the mean position of the $j$-the beamlet on the screen, here determined by the mean value of the Gaussian fit on the beamlet with RMS size of $\sigma_{\mathrm{beamlet}_j}$, and $\overline{x}^{\prime}$ is the mean divergence of all beamlets.

The angular spread contribution exclusively from the emittance of the beam $\sigma_{x^{\prime}_j}^{\prime\,2}$ must be deconvolved in quadrature from the beamlet distribution as~\cite{Sears.2010, Apsimon.2019},
\begin{equation}
    \sigma_{x^{\prime}_j}^{\prime\,2} = \frac{\sigma_{\mathrm{beamlet}_j}^{2} - \left( d M / \sqrt{12}\right)^2}{\Lpps^2} \, ,
    \label{eq. deconvolution term}
\end{equation}
where the term $\left( d M / \sqrt{12}\right)$ is the RMS of the magnified beamlet diameter when considering it being a transversal flat-top distribution, where $M=1+\Lpps/\Ljpp$ is the geometrical magnification constant, with $\Ljpp$ as the distance between the source and the mask and $\Lpps$ as the distance between the mask and scintillation screen (detector position). The term $\sigma_{\mathrm{beamlet}_j}$ is given by the measurement of the RMS size of each beamlet at the scintillation screen during the experiment.

\section{Testing the pepper-pot method}
\label{sec. testing pp method}
Now that we have specified the design parameter for the pepper-pot mask and the procedure for calculating the beam RMS emittance, we use Monte Carlo simulations to benchmark the emittance evaluation, enabling us to later investigate the limitations of the method.

\subsection{GEANT4 simulation setup}
\label{subsec. GEANT4 simulation}
To evaluate the method and identify its limitations (with an emphasis on parameters typical of LWFA beams), we conducted GEANT4 Monte Carlo simulations~\cite{Agostinelli.2003, Allison.2006, Allison.2016} to generate synthetic data that is used to retrieve the emittance and compare it to the input beam. The choice of GEANT4 for simulations is motivated by the need to accurately model the diagnostic response under experimental conditions. This includes not only the electron beam signal at the scintillation screen but also background contributions from scattered particles and X-rays, which can influence both the signal and data analysis. Particle codes, such as PARMELA, GPT, OPAL and ELEGANT~\cite{Poplau.2004, Young.2003, adelmann.2019, Borland.2000Elegant}, are unable to account for secondary particles generated by interactions between electrons and the high-Z material of the pepper-pot mask.

The parameters for the simulation are derived from the conducted experiment such that $\Ljpp = 181~\si{\mm}$ and $\Lpps = 1269~\si{\mm}$. The geometrical magnification of the experimental layout is $M = 1 + 1269/181 \approx 8$. Different virtual screens, used for numerical evaluation only, were placed in the simulation volume to record the parameters of the electrons propagating through them.

A total of $3\times10^6$ primary electrons were generated at the source point, with parameters equivalent to those analyzed in the LWFA experiment and PIC simulation, both discussed later. Consequently, the energy spectrum of the primary beam in the simulations was characterized by a Gaussian distribution, with a mean energy of $72~\si{\MeV}$ and an RMS value of $\sigma_E = 50~\si{\MeV}$, corresponding to an energy spread of approximately $\left( \sigma_E/E \right) \approx 52~\si{\percent}$. Additionally, the RMS divergence of the electron beam was set to $\theta_{x} = 1.85~\si{\mrad}$. These parameters remained fixed throughout all simulations presented here. The emittance of the beam was varied by varying the source size.

The pepper-pot method was tested numerically by calculating the emittance using the parameters obtained from recorded electrons (position, momenta, and energy) at the virtual screen positioned at the same position as an experimental detector screen. The electrons that propagated through the scintillation screen were binned with a resolution of 18~\si{\um}/\si{\pixel}, similar to the experiment. The baseline, visible in the integrated signals, was removed to enhance the signal-to-noise ratio. This background was caused by large-angle scattered particles and X-rays produced via bremsstrahlung on the mask.

Finally, an algorithm for calculating the RMS emittance using Eq.~\eqref{eq. PP equation} was applied to the baseline corrected integrated signals. The emittance was then retrieved through the following steps: (i) identification of peaks representing the centers of each beamlet signal; (ii) fitting of individual Gaussian distributions to each beamlet signal and using their respective values of $n_j$ and $\sigma_{\mathrm{beamlet}j}$; (iii) calculation of the deconvolved beamlet angular spread $\sigma_{x^{\prime}_j}^{\prime}$ using Eq.~\eqref{eq. deconvolution term}; (iv) evaluation of the RMS emittance of the beams with Eq.~\eqref{eq. PP equation}. 

The common approach of fitting a Gaussian distribution~\cite{Sears.2010} presumes that the size of each beamlet on the screen is dominated by its emittance, \textit{i.e.}, that the beamlet size is much larger than the point  projection image of the hole. Note that if the signal at the screen is integrated vertically, \textit{i.e.}, in the y-direction, the emittance evaluated corresponds to the emittance in the x-direction. Similarly, integrating the beamlet signal in the x-direction results in the emittance in the y-direction.

\subsection{Beam emittance for various source sizes}
\label{subsec. beam emittance various sorces}

With the simulation setup and parameters established, multiple simulations were conducted for various source sizes but constant beam divergence, leading to different RMS emittances. Additionally, masks with various hole diameters were employed to evaluate the performance of the pepper-pot method. The RMS emittance of the primary electron beam in the simulations was altered solely by modifying the beam’s source size. The beam’s emittance was then inferred using the algorithm steps described above.

\begin{figure*}[t]
    \includegraphics[width=0.99\linewidth]{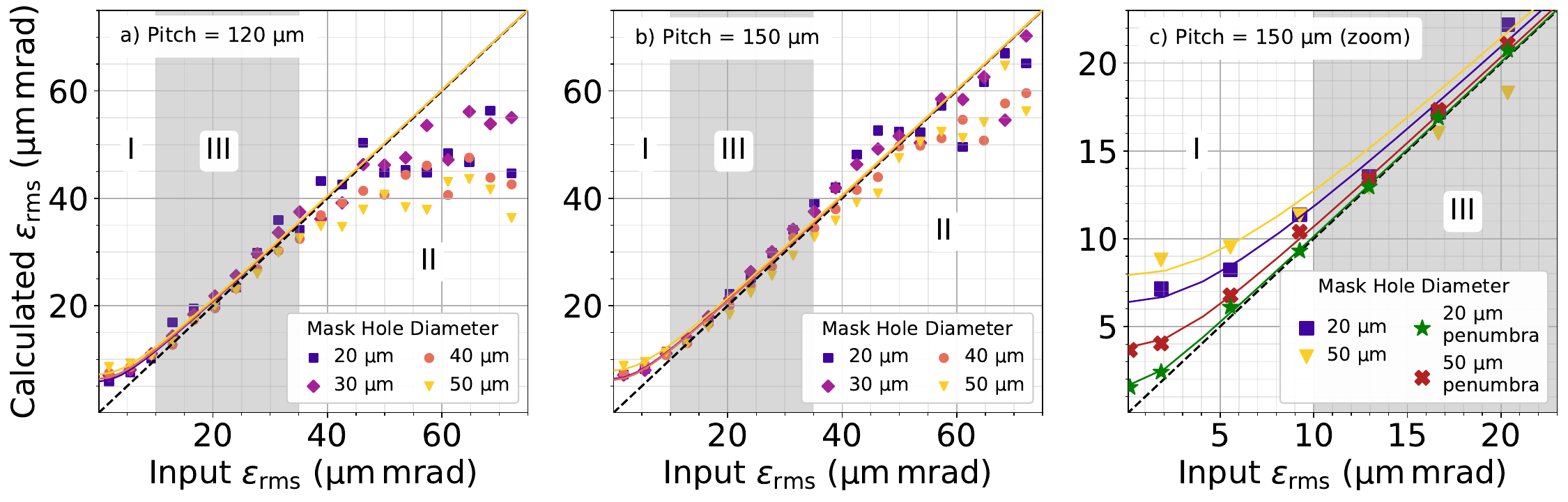}
    \caption{Calculated emittance from GEANT4 simulations for pepper-pot masks with pitches of (a) 120~\si{\um} and (b) 150~\si{\um} across different hole diameters. Panel (c) shows results from the penumbral deconvolution method for the 150~\si{\um} pitch at low emittance. Uncertainties were below 5\% and thus omitted. Curves following $\epsilon_{\rmscalc}^2 = \Omega^2\theta_{x}^2 + \epsilon_{\rmsin}^2$ were fitted to the data, with the black dashed line marking the diagonal. The fitted values of $\Omega$ are shown in Appendix~\ref{appendix:fitting}. Additionally, region I marks where the emittance is overestimated, region II where it is underestimated, and region III where it is recovered with less than 10\% error from the input RMS emittance.
    }
    \label{fig:source scan geant4} 
\end{figure*}

The simulation results for two different geometries are shown in Fig.~\ref{fig:source scan geant4}. The left panel compares the input emittance of the primary electron beam in GEANT4, $\epsilon_{\rmsin}$, with the calculated emittance $\epsilon_{\rmscalc}$ obtained from the simulated beamlet projections on the virtual screen using the algorithm in Section~\ref{subsec. GEANT4 simulation}, for a pepper-pot mask with a 120~\si{\um} hole pitch with hole diameters of $20–50$~\si{\um} in steps of 10~\si{\um}. The center panel shows the same comparison for a mask with a 150~\si{\um} hole pitch. %

The calculated emittance values can be fitted well by using the following model function: 
\begin{gather}
    \epsilon_{\rmscalc}^2 = \Omega^2\theta_{x}^2 + \epsilon_{\rmsin}^2 \approx (\Omega^2 + \sigma_{x}^2)\theta_{x}^2\, ,
    \label{eq. fitting curve source size scan}
\end{gather}
depicted also in Fig.~\ref{fig:source scan geant4}. Here, $\epsilon_{\rmsin} \approx\theta_{x}\sigma_{x}$, where the divergence $\theta_{x}$ is given in \si{\mrad} and the source size $\sigma_{x}$ in \si{\um} as in the emittance definition in Eq.~\eqref{eq. rms emittance definition}. The constant parameter $\Omega$ depends only on the geometric parameters of the experiment, as we will examine in the next section. The best fit values of the $\Omega$ parameters for the different curves in Fig.~\ref{fig:source scan geant4} are provided in Appendix~\ref{appendix:fitting}. 

From the simulation results, three regions—labeled I, II, and III—can be identified in each plot in Fig.~\ref{fig:source scan geant4}. Region I occurs at low emittance values ($< 10~\si{\um\mrad}$), where the beam emittance is no longer resolved, leading to an overestimation of the true value. Region II appears at high emittance values ($> 35~\si{\um\mrad}$), where the emittance is underestimated. Region III lies between these limits, where the source size is accurately retrieved. These regions are discussed in detail in the following sections.

\section{Limitations of the pepper-pot method}
\label{section:limitation of pp method}
Based on the benchmark simulations presented previously in Section~\ref{sec. testing pp method}, this section investigates the limitations of the pepper-pot technique. The simulation results in Fig.~\ref{fig:source scan geant4} can divided into three different regions, each of which is analyzed in detail in the following subsections.
\subsection{Challenges of resolving low emittance values}
\label{subsection problem resolving low emittances}
\begin{figure}
    \centering
    \includegraphics[width=\linewidth]{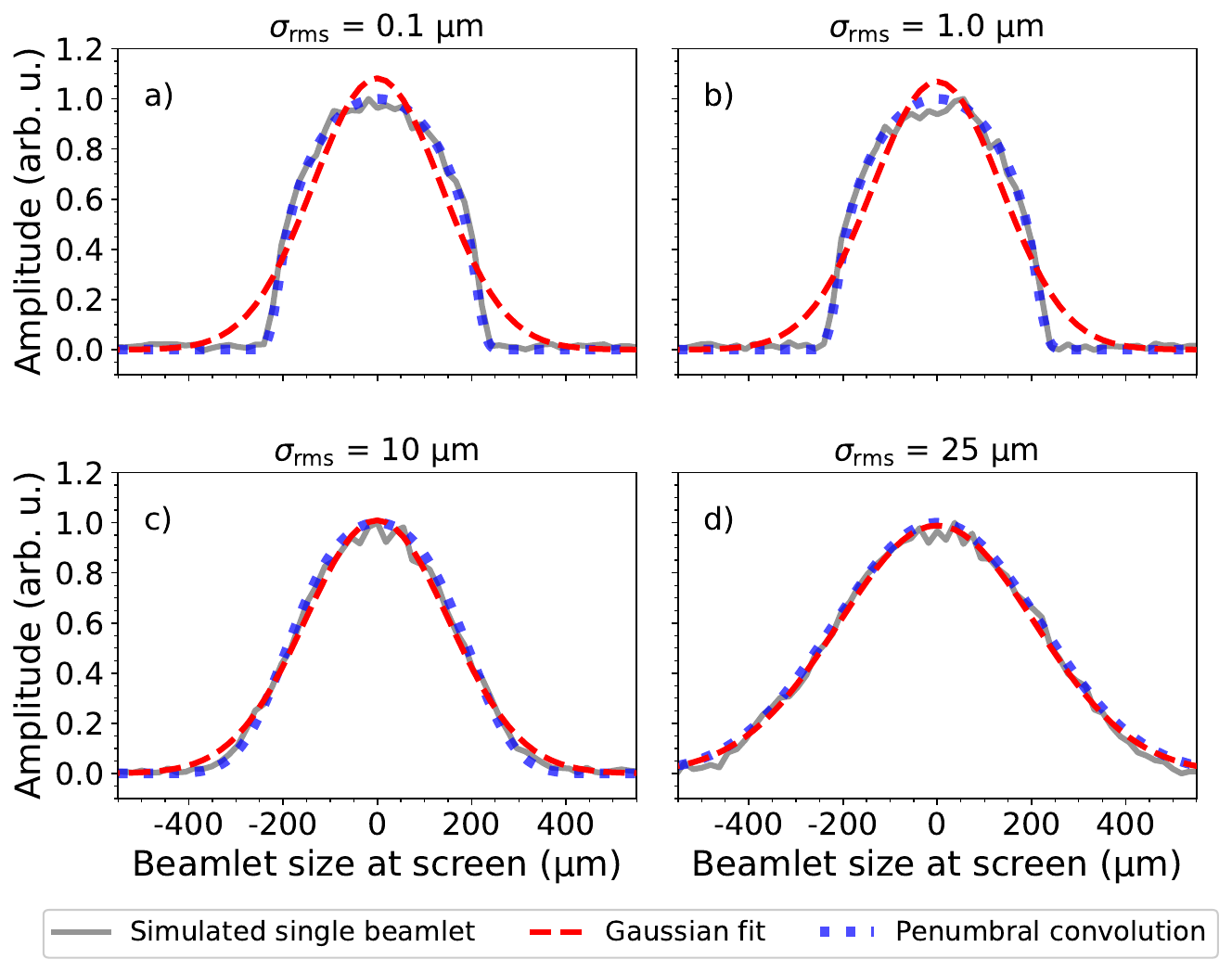}
    \caption{
    Comparison of the fit model for the beamlets depending on the source size of the electron beam.
    Gaussian fit (red dashed line) and penumbral convolution (blue dotted line) between a Gaussian source and the mask holes fitted to the simulated beamlet image (solid gray line) for various beam waists $\sigma_\mathrm{rms}$ while maintaining a fixed beam divergence of $1.85$~\si{\mrad}. 
    }
    \label{fig. gauss fit in beamlet image}
\end{figure}

In the first region of interest of the plots of Fig.~\ref{fig:source scan geant4}, region I, the angular spread of the beamlets is dominated by the point projection of the mask´s holes. The finite source size is visible in the penumbral broadening of the individual beamlets on the screen as seen in Fig.~\ref{fig. gauss fit in beamlet image}. Consequently, the Gaussian distributions on each beamlet are a poorly matched fit to the beamlet shape. The data analysis method previously mentioned in section~\ref{subsec:evaluating beam emittance}, then incorrectly assumes the angular spread $\sigma_{\mathrm{beamlet}}$ to be due to emittance only and returns a value comparable in magnitude to the term $\left( d M / \sqrt{12}\right)$ given in Eq.~\eqref{eq. deconvolution term}. 
Hence, the angular spread inferred by the Gaussian fit tends towards a constant (sometimes even a negative) value. With an increase in beam waist, the Gaussian fit returns a good estimate of the beamlet angular spread, as  the spread due to emittance now exceeds the point projection contribution. 
This issue is illustrated in Fig.~\ref{fig. gauss fit in beamlet image}, where various Gaussian fits on beamlets from different source sizes but with fixed divergence are shown. For small source sizes, \textit{i.e.}, $\sigma_{x} < 10~\si{\um}$, the Gaussian fit does not represent the data well, and the value of the deconvolved angular spread in this range is $\approx 60~\si{\urad}$. 

In the absence of scattering, the integrated shape of beamlet $i(x)$, as imaged during the experiment, can be accurately described by an integral over the transverse axis (here, the y-direction) of the two-dimensional convolution of the source with the hole and the imaging system, as described in~\cite{Chen.1998, Haykin.1994}:
\begin{equation}
    i(x)= \int h(x,\,y) \ast s(x,\,y) \ast \mathrm{PSF}(x,\,y) \, \mathrm{d}y \,,
    \label{eq. penumbral convolution}
\end{equation}
where $s(x,\,y)$ represents the source distribution before the PP mask, assumed to follow a Gaussian profile; $h(x,\,y)$ describes the hole geometry; and $\mathrm{PSF}(x,\,y)$ is the point spread function (PSF) of the imaging system.

One can deconvolve the signal $i(x)$ to obtain $s(x,\,y)$ using, for example, Wiener filters~\cite{Chen.1998}, or by solving a minimization problem to fit the best source size, assuming that the source function follows a Gaussian distribution. We opted for the latter method, which we will demonstrate later.

In our case, the function $i(x)$ is obtained by integrating the signals of the beamlets on the screen. The function $h(x,\,y)$ is assumed to be a top-hat function representing the magnified hole of the mask, and the PSF is determined through calibration of the imaging system. Here, we assume a Gaussian-like PSF with an RMS value of $\sigma_\mathrm{psf} \approx 6~\si{\um}$.

The results of the analysis performed for source sizes of 0.1~\si{\um} and 1.0~\si{\um} are shown in Fig.~\ref{fig. gauss fit in beamlet image}~a) and b). From these results, we observe that the convolution provides a better fit instead of a Gaussian function. Consequently, the inferred source sizes $\sigma_x$ for these small emittance beams are also reduced, as seen in Fig.~\ref{fig:source scan geant4}~c).

When analyzing the GEANT4 simulation results using the alternative method of the penumbral convolution, the constant parameter $\Omega$ given in Eq.~\eqref{eq. fitting curve source size scan} reduces from values between $3.3 - 4.5~\si{\um}$ to smaller values: $\Omega \approx 2.0~\si{\um}$ for the pepper-pot with a hole diameter of 50~\si{\um}, and $\approx 0.9~\si{\um}$ for the pepper-pot with a hole diameter of 20~\si{\um}. Therefore, despite reducing the calculated emittance obtained by applying the deconvolution method to the penumbra of the beamlet projected onto the screen, the pepper-pot method still fails to resolve very small emittance values.

Hence, as the beam emittance of the input electron beam decreases such that $\epsilon_{\rmsin} \rightarrow 0$, the minimum calculable RMS emittance approaches $\epsilon_{\rmscalc} \approx \Omega\,\theta_{x}$, where $\Omega$ can be obtained from Eq.~\eqref{eq. penumbral convolution} as
\begin{equation}
\Omega^2 = \frac{1}{12} \left( \frac{\Ljpp}{\Lpps} \right)^2 \left( d^2 + \sigma_\mathrm{psf}^2 \right) \, ,
\label{eq.alpha_limiting_factor}
\end{equation}
which is a constant determined solely by the geometric parameters of the experiment. The factor $1/12$ arises from the RMS width of a top-hat function representing the hole and the PSF. For geometric magnifications $M \gg 1$ and $d \gg \sigma_\mathrm{psf}$, Eq.~\eqref{eq.alpha_limiting_factor} can be approximated as $\Omega\approx d / \left( \sqrt{12} M \right)$.

For the GEANT4 simulation parameters used in this work, the minimum achievable $\epsilon_{\rmscalc}$ with a mask of 150~\si{\um} pitch is approximately 1.6~\si{\um\mrad} for a hole diameter of 20~\si{\um}, and 3.8~\si{\um\mrad} for holes of 50~\si{\um} diameter. These values agree with those obtained from the fit functions of the deconvolved data using the penumbral method as shown in Appendix~\ref{appendix:fitting}.

To mitigate the limitations of the pepper-pot method in resolving small emittance values, several options can be considered. For example, one could reduce the pitch distance or the diameter of the mask holes. However, by reducing the diameter of the holes, the number of particles propagating through them is reduced, making it challenging to achieve a signal-to-noise ratio at the screen much larger than one. In addition, this presents challenges with the current technology available for machining tiny holes in pepper-pot masks. The common method involves laser drilling, which is limited by the thickness of the mask in terms of the hole diameters it can produce. If smaller holes are required, the thickness of the mask must also be reduced. However, reducing the mask thickness results in fewer electrons being scattered at larger angles or being stopped, leading to a larger background noise that can reduce the accuracy of the emittance measurement.

Another possible solution is to increase the setup geometry distances so that the ratio $\left( \Ljpp/\Lpps \right)$ is small enough to minimize the parameter $\Omega$, regardless of the pepper-pot hole size and the imaging system response. However, in doing so, the pepper-pot mask cannot be placed too close to the LWFA target to prevent damage from the laser beam, and the diagnostic screen should not be positioned too far from the pepper-pot mask to ensure a high signal-to-noise ratio of the beamlets. This results in an excessively long setup, making it impractical due to the large distances involved.

For example, for our electron beam with a divergence of 1.85~mrad, measuring emittances below 1~\si{\um\mrad}, \textit{i.e.}, a constant of $\Omega < 0.54~\si{\um}$—would require, assuming the pepper-pot mask design and imaging system characteristics remain unchanged and the distance $\Ljpp$ is fixed at about 180~mm to avoid laser damage to the pepper-pot mask, a minimum drift length of 4873~\si{\mm}. This large drift length, required to detect such low emittance, defeats the purpose of using the pepper-pot for compact diagnostics. Additionally, the signal-to-noise ratio, assuming a constant beam charge, would be reduced by a factor of approximately 12, making it more difficult to detect the beamlets on the scintillating screen.

\subsection{Underestimating large emittance values}
\label{subsec: understimating large emittance values}
\begin{figure}
    \includegraphics[width=0.9\linewidth]{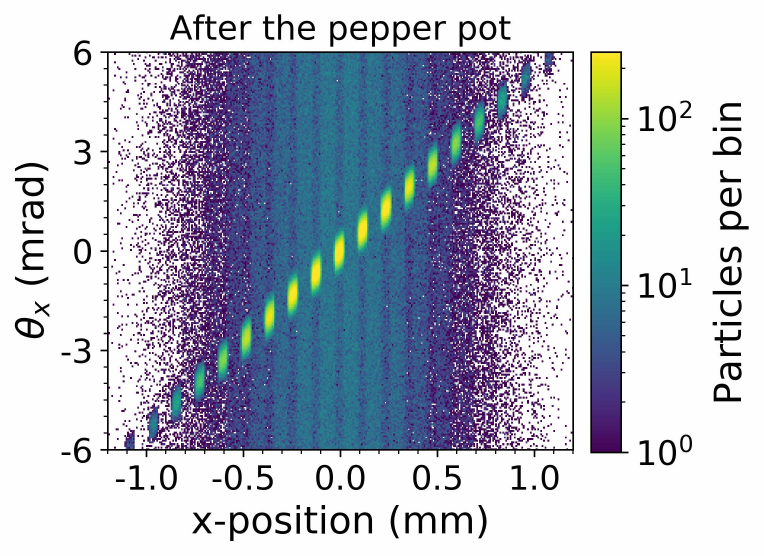}
    \caption{Simulated trace space of an electron beam with a 40~\si{\um} source size and a beam emittance of 74~\si{\um\mrad} (from Fig.~\ref{fig:source scan geant4}~a)) after its interaction with a pepper-pot mask with $d=50$~\si{\um} and $\Lambda=120~\si{\um}$. The trace space of the beam is sampled in beamlets, and background noise arises due to electrons that are scattered at large angles. 
    }
    \label{fig. trace space large source sizes}
\end{figure}

For values above $>35~\si{\um\mrad}$, the emittance values are underestimated for large source sizes, as identified as region II of the plots in Fig.~\ref{fig:source scan geant4}. The reason why the method fails to infer correctly the emittance values in this region is better understood by examining the trace space of the beam after its interaction with the mask and at the detection plane, as shown in Fig.~\ref{fig. trace space large source sizes}.

The sampling of the beam into beamlets in the pepper-pot mask is represented by the trace space shown in Fig.~\ref{fig. trace space large source sizes}. As the beamlets propagate from the mask to the detection screen, the trace space of each individual beamlet rotates, and the beamlets rapidly spread due to their angular divergence. As the overlap intensifies, the baseline of the integrated signal (in the angular direction) also elevates, since the beamlets now start to overlap each other, diminishing the signal-to-noise ratio of the individual beamlets during post-processing of the data. The beamlet overlap can be minimized during data evaluation through the implementation of baseline correction. However, applying baseline reduction during data analysis also reduces the height $n_j$ and RMS spread $\sigma_{\mathrm{beamlet}_j}$ of the beamlets, resulting in an underestimation of the calculated emittance using Eq.~\eqref{eq. PP equation}.

\begin{figure}
    \centering
    \includegraphics[width=0.99\linewidth]{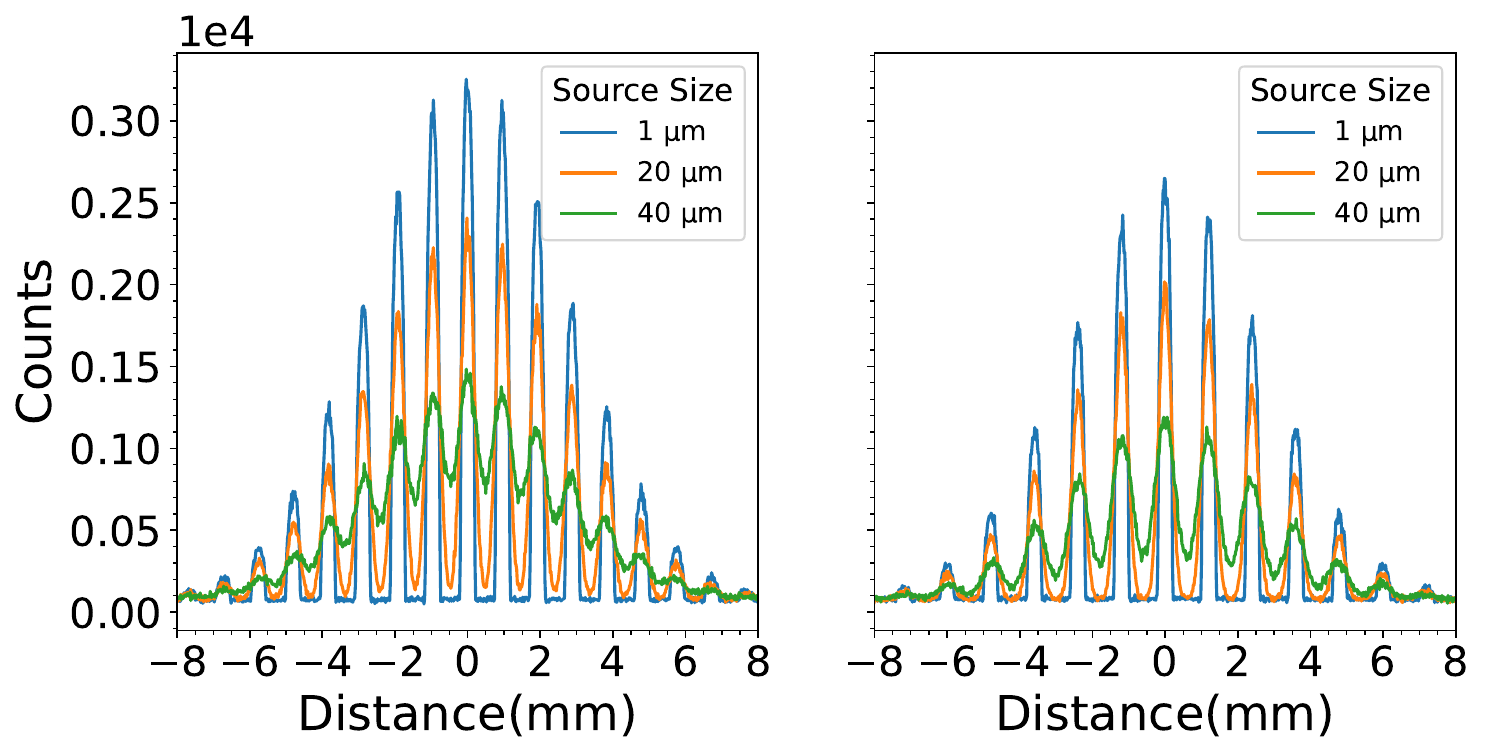}
    \caption{Integrated signals (summation over the y-axis to investigate $\epsilon_{x}$) from GEANT4 simulation results of the beamlet signal at the screen for a pepper-pot mask with a hole diameter of 50~\si{\um} and a $\Lambda=120$~\si{\um} (left) and 150~\si{\um} (right). The increased pitch size $\Lambda$ leads to a lower overall baseline.}
    \label{fig: geant4 integrated signals}
\end{figure}

To effectively mitigate these overlapping effects, a larger distance between the pepper-pot holes $\Lambda$ can be utilized. A comparison between the integrated signal obtained from different masks with pitches of 120~\si{\um} and 150~\si{\um} is shown in Fig.~\ref{fig: geant4 integrated signals}. For the simulations employing a large pitch distance in the pepper-pot, the beamlets for large source sizes have a reduced overlap (see the right panel of Fig.~\ref{fig: geant4 integrated signals}), contrasting with the integrated signals of the smaller pitch pepper-pot mask shown in the left panel of the same figure. This improvement is also noticeable in the results depicted in panels~b) and c) of Fig.~\ref{fig:source scan geant4}. Increasing the pitch distance of the pepper-pot mask reveals that the region where overlap occurs is situated at higher emittance values, in our case above 35~\si{\um\mrad}. Consequently, the method becomes capable of characterizing a broader range of beam emittances. This adjustment in the pepper-pot mask design enables the method to diagnose beams with large emittances and source sizes.

To determine the minimum pitch distance to avoid overlap on the screen at the $1/e^2$ level of the angular spread of the beamlets, a modified version of the condition given in Ref.~\cite{Delerue.2009} can be used: 
\begin{equation} 
\Lambda > \sqrt{2} \left[ 2 \sigma_\mathrm{rms} \left( 1 - \frac{\Ljpp}{\Lpps}\right) + d \right] \,, 
\label{eq. condition pitch} 
\end{equation} 
where $d$ is the hole diameter of the mask, and $\sigma_{\rms}$ is the source size of the electron beam. The condition given in Eq.~\eqref{eq. condition pitch} depends on the distance between the electron beam source and the mask, $\Ljpp$, and the screen distance from the pepper-pot, $\Lpps$. For larger distances of $\Lpps$, the pitch between the mask holes should also be increased. The $\sqrt{2}$ term arises from the conversion of a top-hat function to the beam waist of a Gaussian.

To verify this condition, it can be applied to the simulation data shown in Fig.\ref{fig:source scan geant4}. For example, for an input GEANT4 RMS emittance of 40~\si{\um\mrad} and a mask with holes of diameter $d = 50~\si{\um}$, the minimum pitch required for imaging the beamlets on the screen (for the simulated setup) is $\Lambda \geq 120~\si{\um}$. By analyzing the simulation results, we observe that with a pitch of 120~\si{\um}, the mask with 50~\si{\um} holes starts to fail in resolving emittances around 40~\si{\um\mrad} as predicted by Eq.~\eqref{eq. condition pitch}, corresponding to a source size of $\approx 22~\si{\um}$. On the other hand, when the pitch is increased to 150~\si{\um}, the pepper-pot with the same hole diameter successfully resolves the emittance of the electron beam with the same source size.

\subsection{Pepper-pot operating regime}
\label{section:inferring emittance with pp method}

For values in the range between 10 to 35~\si{\um\mrad}, the pepper-pot method accurately resolves the beam emittance for our conditions. This range is shown as region III in Fig.~\ref{fig:source scan geant4}. In contrast to the other limiting regions described previously, here the overlap between the beamlets is minimal, enabling the correct retrieval of their individual spreads and heights after the removal of background noise (baseline).

For these intermediate emittance values, the Gaussian fit on the integrated beamlet signals can well represent the integrated signal of the beamlets on the virtual screen at the detector plane, as seen in Figs.~\ref{fig. gauss fit in beamlet image}~c) and d) allowing the emittance to be calculated using the method described in section~\ref{subsec:evaluating beam emittance}.

The application regime of the mask is identified by the angular spread of the beamlets and the source sizes being significant enough that the second term on the right-hand side of Eq.~\eqref{eq. fitting curve source size scan} predominates over the constant term $\Omega$. Consequently, the dependence on the experimental layout of the setup is minimized such that $\Omega\theta_x \ll \epsilon_{\rmsin}$, and the emittance is approximated by the beam parameters as $\epsilon_{\rmscalc} \approx \theta_{x}\sigma_{x}$.

\section{Emittance measurement of electron bunches of an LWFA}
\label{section:experiment}

\subsection{Experimental setup}
\begin{figure*}
	\centering
	\includegraphics[width=0.95\linewidth]{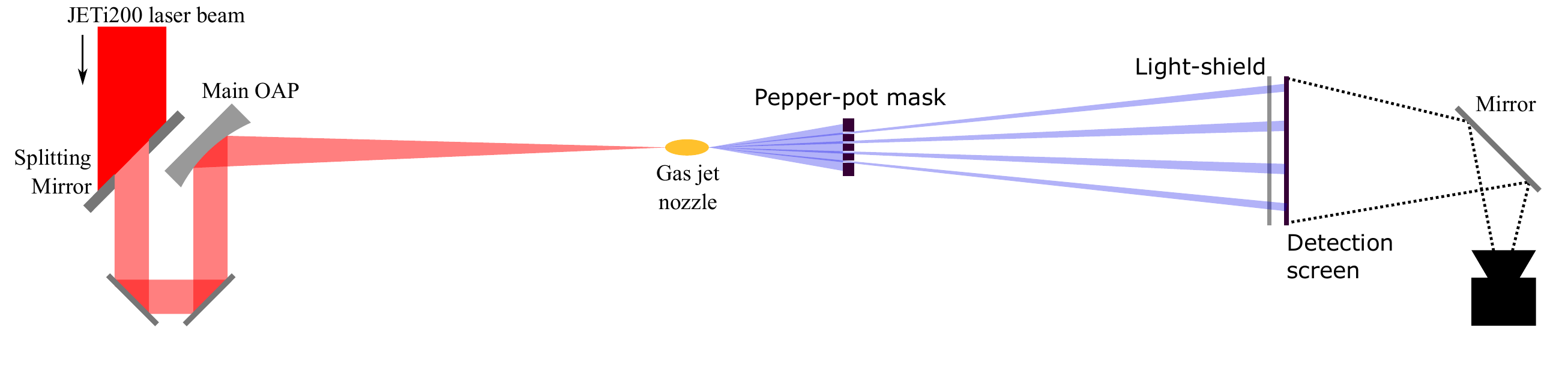}
	\caption{Experimental layout for emittance measurements using the pepper-pot mask.}
	\label{fig. experimental setup up}
\end{figure*}

The JETi200 laser system provides laser pulses with an energy of 4.6~\si{\J} centered at 800~\si{\nm}, and a pulse duration of 23~\si{\fs}. The experimental setup is illustrated in Fig.~\ref{fig. experimental setup up}. The layout of the experiment was similar to the geometry used in the Monte Carlo simulations discussed earlier.

In the experiment, the beam was reduced from its original size of 120~mm to a smaller diameter of 60~mm by a splitting mirror resulting in a total beam energy of 1.15~J. The laser beam was focused by an off-axis parabolic mirror (f-number~=~16.7) to a spot with $(23.7 \pm 1.8)$~\si{\um} at full width at half maximum (FWHM) with approximately 38\% of the pulse energy within the FHWM, resulting in a peak intensity of $7\times 10^{18}$~\si{\W\per\cm\squared}. The focused laser beam impinged on a supersonic gas jet (mixture of 95\% He and 5\% N$_2$) generating a plasma with an electron density of $1.1\times10^{19}$~\si{\cm^{-3}}.

In the experiment, the distance between the exit of the gas jet nozzle and the mask was about $\Ljpp = (180 \pm 1)~\si{\mm}$. Finally the distance from the mask to the YAG:Ce scintillating screen was approximately $\Lpps = (1269 \pm 1)$~\si{\mm}. The geometrical magnification of the setup was $M = (\Ljpp + \Lpps)/\Ljpp \approx 8$.

The pepper-pot used in this experiment has a square grid of 33~$\times$~33 holes of $d \approx 50$~\si{\um} diameter with a pitch length of $\Lambda=120$~\si{\um}. The mask is $200$~\si{\um} thick and made of tungsten (density of 19.3~g/cm$^3$ and collision stopping power of about 1.4~\si{\MeV\cm\squared\per\gram} for 73~\si{\MeV} electrons~\cite{NIST.2022}).

As the beam propagates through the mask, the transmitted beamlets travel towards a YAG:Ce scintillation screen (detection plane) of $100~\si{\nm}$ thickness, emitting a broad spectrum with a peak wavelength at $\lambda \approx 550~\si{\nm}$, and imaged by an Andor Marana camera (16-bit, quantum efficiency $\approx 95\%$ at $550~\si{\nm}$)~\cite{OxfordInstruments.2022} with an optical resolution of $r_i \approx 18.5~\si{\um}/\si{pixel}$. To suppress background noise in the electron signal caused by scattered laser light, an Aluminium foil of approximately $100~\si{\um}$ thickness was placed about $10~\si{\mm}$ in front of the screen to light-shield the diagnostic.

\subsection{Electron beam characteristics}
The LWFA electron beam had a maximum energy of about 120~MeV with a total charge of approximately $(5.6\,\pm\,0.7)$~pC. The electron energy distribution has a weighted average of approximately 73~\si{\MeV}, equivalent to $\langle \gamma \rangle \approx 143$, and a weighted average energy spread of $\left( \sigma_E/E \right) = (27.3 \pm 4.8)\%$. 

The RMS divergence of the beam in the x-direction obtained during the experiment was evaluated to be $\theta_{x} \approx (1.8\pm0.3)$~\si{\mrad}. To determine its value, the pepper-pot mask was removed from he beam path and recorded the beam profile on the detection screen. For each captured beam profile, a two-dimensional Gaussian was fitted to obtain their respective RMS spread. Finally, the beam divergence was calculated by averaging all evaluated spreads, with the error determined by their standard deviation.

For more information on the electron beam, see Ref.~\cite{Salgado.2024}.

\subsection{Emittance measurement using pepper-pot mask}

\begin{figure}
	\centering
	\includegraphics[width=0.9\linewidth]{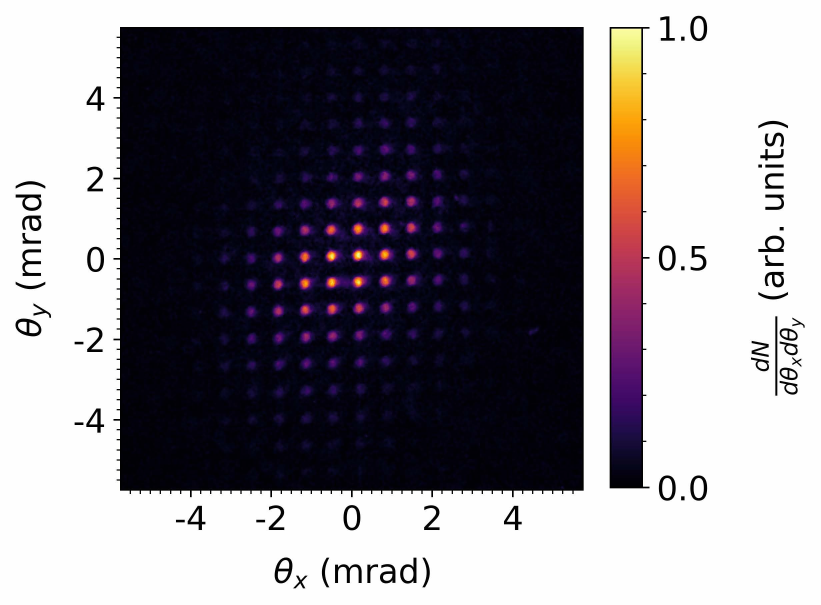}
	\caption{Example of beamlets imaged using a scintillation screen. The coordinates $\theta_x$, $\theta_y$ represent the beam divergence in the horizontal and vertical directions, respectively.}
	\label{fig. pepper-pot series of shots}
\end{figure}

Figure~\ref{fig. pepper-pot series of shots} shows an example of pepper-pot beamlets imaged on the YAG:Ce screen. From this beamlet images, an average distance between beamlets of $(953.6\pm~15.2)$~\si{\um} was obtained. To calculate the emittance using Eq.~\eqref{eq. PP equation}, we used the same post-processing chain and algorithms described previously in Section~\ref{sec. testing pp method}. 

After analyzing 100 shots using the standard evaluation method, which uses Eq.~\eqref{eq. PP equation}, we measure an average geometrical emittance of $\epsilon_{\rmsexp}= (26.2\pm7.3)~\si{\um\mrad}$ for the pepper-pot mask with 50~\si{\um} hole size and 120~\si{\um} pitch, where the uncertainty is given by the standard deviation of all analyzed shots.

\begin{figure}
	\centering
	\includegraphics[width=0.85\linewidth]{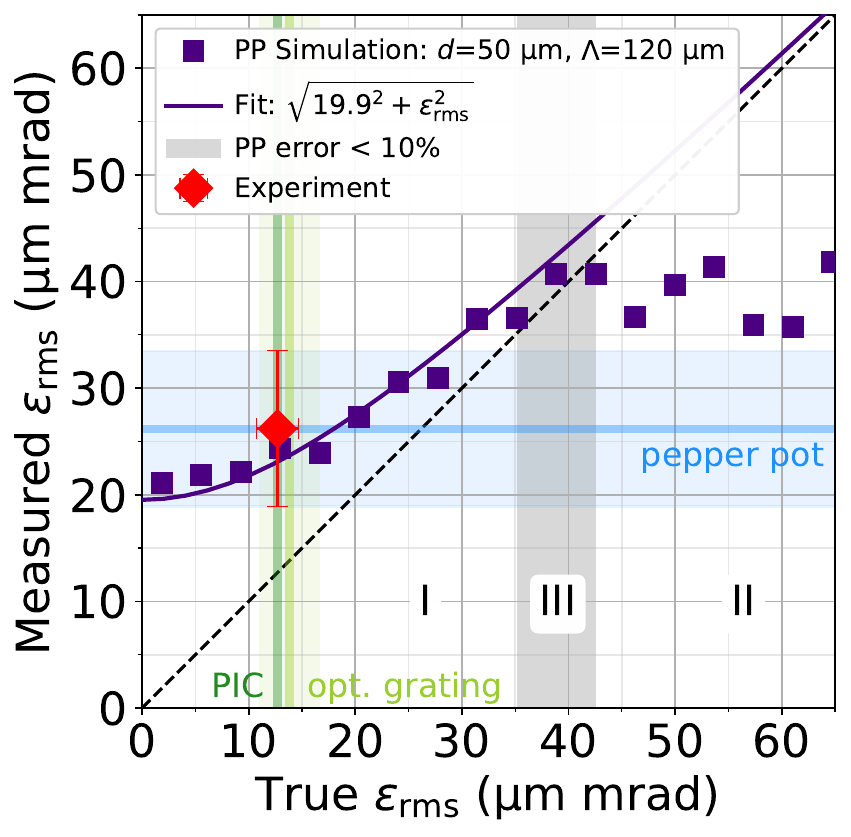}
	\caption{
	Comparison of experimental and simulated emittance for our setup. The PP measurement yields $(26.2 \pm 7.3)~\si{\um\mrad}$ (blue region, red diamond). GEANT4 simulations of the PP diagnostic (purple symbols and fit) indicate that the measurement lies in region I, where the PP method overestimates the true emittance. Region III (gray) denotes the range in which the PP setup achieves better than 10\% accuracy. The green vertical shaded regions represent the expected beam emittance from PIC simulations and from the laser grating diagnostic of Ref.~\cite{Salgado.2024}. Using these expected (true) values in the GEANT4 simulation of the PP diagnostic yields results consistent with the PP measurement. 
	}
	\label{fig:geant4_experiment}
\end{figure}

By comparing our measurement with GEANT4 Monte Carlo simulations of the pepper-pot diagnostic for our experimental setup (purple symbols and fit curve in Fig.~\ref{fig:geant4_experiment}), we determine that our measurement is located in region I (as defined in Section~\ref{subsec. beam emittance various sorces}, see also Fig.~\ref{fig:source scan geant4}), where the pepper-pot method overestimates the true beam emittance. From this alone, we can only determine that the true beam $\epsilon_{\rms}$ is smaller than approximately 30~\si{\um\mrad}.

Moreover, Fig.~\ref{fig:geant4_experiment} shows that our PP setup can resolve emittances with less than 10\% error only within a narrow window between 35 and 42~\si{\um\mrad}, highlighted by the gray shaded area (region III).

The PIC simulations performed with the same experimental parameters (see Appendix~\ref{appendix:pic}) yield an expected RMS beam emittance of about $12.7~\si{\um\mrad}$ (dark green shaded region in Fig.~\ref{fig:geant4_experiment}). In addition, we have an independent experimental inference of the beam emittance for the same LWFA setup using the laser grating method~\cite{Seidel.2021} of 
$(13.8\pm2.8)~\si{\um\mrad}$~\cite{Salgado.2024} (light green shaded region in Fig.~\ref{fig:geant4_experiment}), which is very close to the PIC simulation result. Taking those values as input for the GEANT4 simulation of the diagnostic, the emittance $\epsilon_{\rmscalc}$ determined from the synthetic diagnostic agrees with the experimentally measured value $\epsilon_{\rmsexp}$ within the experimental uncertainty (red diamond).

We conclude that the limitations of the pepper-pot method prevented a proper determination of the true beam emittance in this case. 
However, our GEANT4 simulations of the diagnostic establish a consistent picture with those emittance values obtained via better-suited low-emittance diagnostic methods, and PIC simulation results.

For completeness, the averaged normalized emittance was evaluated using Eq.~\eqref{eq. norm-emittance with drift} for the experimental data results in $\epsilon_n \approx 158.8~\pi~\si{\mm\mrad}$. The large normalized emittance value arises due to the large energy spread of the electron beam, resulting in different phase space rotation speeds of the particles while freely drifting toward the screen. This large value can also be understood by examining Eq.~\eqref{eq. norm-emittance with drift}. Due to the substantial energy spread and drift length, the first term on the right-hand side is much larger than the RMS emittance term, $\left( \sigma_E/E \right)^2 \theta_{x}^4 L \gg \epsilon_{\rms}^2$. Consequently, the normalized emittance is dominated by the growth of the transverse distribution of the beam.

Our normalized emittance is found to be larger than the values reported in the literature for LWFA beams, such as in Refs.~\cite{Fritzler.2004, Shanks.2009, Sears.2010, Brunetti.2010, Manahan.2014}. The discrepancy arises because we calculate the normalized emittance taking into account the broadband energy spectrum of our electron beam, a consideration that is not accounted for in the cited literature, which considers their energy spread to be negligible as found in RF accelerators.

\section{Conclusions}
\label{section:conclusions}

The pepper-pot method has been used to measure the RMS emittance of electron beams, however, it has limitations that affect its accuracy, particularly when resolving very low emittance values from electron beams generated by LWFA accelerators. In the low emittance regime, below $10~\si{\um\mrad}$, the method struggles due to the dominance of the point projection image over the actual angular spread of the beamlets. This leads to an overestimation of the emittance when applying Eq.~\eqref{eq. PP equation} with input values based on fitted Gaussian distributions.
One opportunity to extend the accuracy of the pepper-pot technique for small source sizes relies in using penumbral deconvolution techniques.
While this improves the accuracy of the calculated and measured source sizes for small emittance beams, it still does not allow for accurate resolution of the emittance due to the small effects of the source size on the overall beamlet size.

For large emittance values, above $35~\si{\um\mrad}$ for our parameters, the method underestimates the emittance due to significant beamlet overlap and the resulting elevated baseline of the integrated signal. Adjustments such as increasing the pitch distance between the holes in the pepper-pot mask help to mitigate these effects, enabling the method to better characterize beams with large emittance.

In the intermediate range (from 10 to 35~\si{\um\mrad} for our parameters), the pepper-pot method accurately resolves the beam emittance. Here, the beamlets' angular spread and source sizes are sufficiently large, reducing the influence of the experimental setup's geometric parameters and allowing reliable emittance measurements.

An experiment performed at the JETi200 laser system to measure the emittance of laser-accelerated electron beams using the pepper-pot method demonstrated the limitations of this technique. The measured value of the RMS emittance of $(26.2\pm7.3)~\si{\um\mrad}$ is larger than expected for LWFA accelerators, but it is consistent with the limited resolution predicted by Monte Carlo simulations for the pepper-pot diagnostic. The input beam emittance for these simulations agrees with the RMS emittance of $12.7~\si{\um\mrad}$ obtained from PIC simulations of the same laser–plasma accelerator, as well as with an independent emittance measurement of $(13.8\pm2.8)~\si{\um\mrad}$ previously reported using the laser grating method~\cite{Salgado.2024}.

In conclusion, for typical LWFA experiments with emittances below tens of~\si{\um\mrad}, the pepper-pot technique cannot provide accurate measurements of small emittances. This method works better for larger emittances, where one can obtain an accurate emittance measurement by choosing an appropriate hole diameter and pitch.
However, for the smallest emittances on the order of tens of picometers~\cite{Hidding.2012}, other techniques like quadrupole scans or the laser-grating method are more appropriate~\cite{Seidel.2021, Salgado.2024}. Since each pepper-pot configuration can only measure a limited range of values accurately, different geometries are needed depending on the actual emittance.

\section{Acknowledgements}

The authors thank G.~Schäfer for operating the JETi200-laser system. This research was funded by the Federal Ministry of Education and Research of Germany (BMBF) in the Verbundforschungsframework (Project Nos. 05K19SJA, 05K22SJB and 05K22SJA). F.~C.~S. and M.~Z. thank the funding by the Deutsche Forschungsgemeinschaft (DFG) under Project No.~416708866 within the Research Unit FOR2783. The research leading to the presented results received additional funding from the European Regional Development Fund and the State of Thuringia via Thüringer Aufbaubank TAB (Contract Nos.~2019~FGI~0013, 2022~FGI~0003, 2022~FGI~0005 and 2023~FGI~0023).

\appendix

\section{Particle-in-cell simulations results}
\label{appendix:pic}
To validate the approximations in Eq.~\eqref{eq. rms emittance definition}, where the cross-correlation term in the RMS emittance definition was neglected, particle-in-cell (PIC) simulations were performed using FBPIC~\cite{Lehe.2016} with parameters representative of the experiment. The plasma electron density followed a third-order super-Gaussian profile with $\approx2.5$~\si{\mm} plateau and a peak density of $10^{19}$~\si{\cm^{-3}}, formed by a gas mixture of 95\% He and 5\% N$_2$. The laser pulse had a peak normalized vector potential of $a_0 \approx 1.6$.

\begin{figure}
    \includegraphics[width=0.99\linewidth]{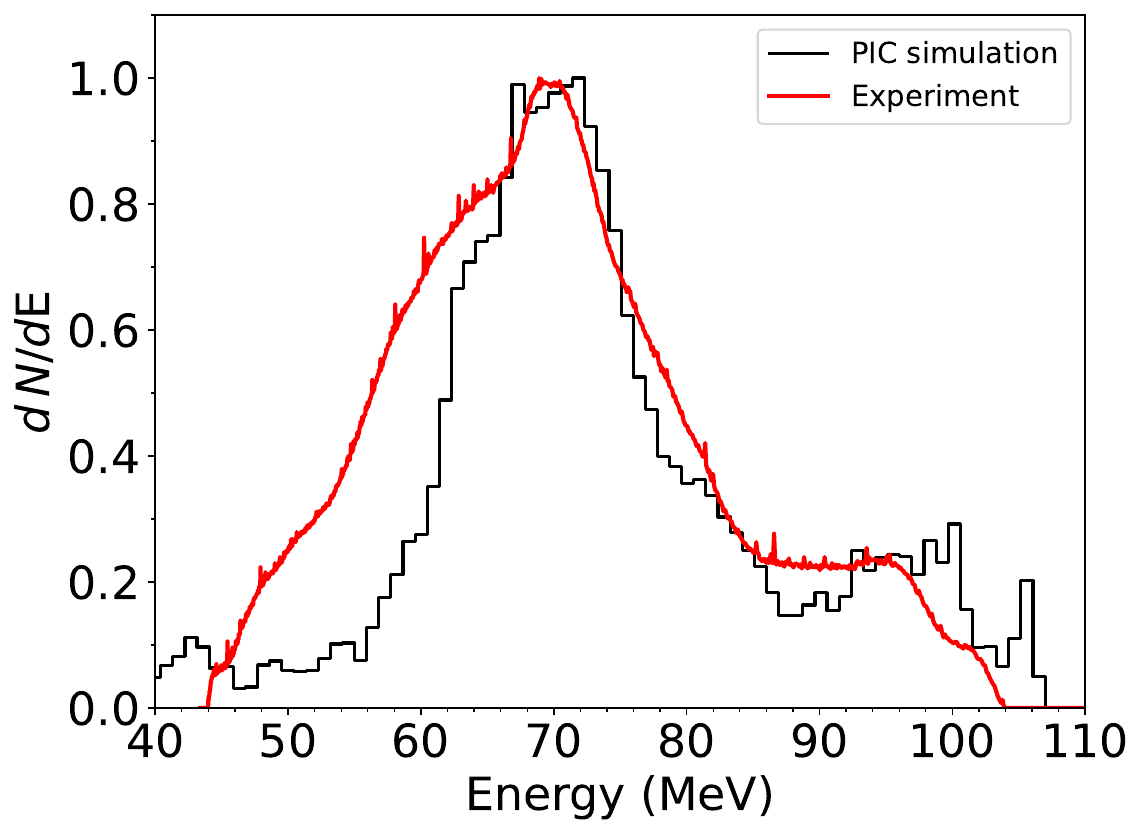}
    \caption{Comparison of electron energy spectra from PIC simulations and experiment. The simulations used laser and plasma density parameters matching those in the experiment. The simulated beam shows an average energy of $(72\,\pm\,23)$~\si{\MeV}, consistent with the experimental value of $(73\,\pm\,19)$~\si{\MeV} reported in Ref.~\cite{Salgado.2024} for the same laser-plasma accelerator.}
    \label{fig-sims-spectrum-comp}
\end{figure}

The simulations reproduced an electron energy spectrum consistent with Fig.~\ref{fig-sims-spectrum-comp}, confirming ionization-induced injection. The simulated beam had an average energy of $(72\pm23)$~\si{\MeV}, in good agreement with the experimentally measured mean energy of 73~\si{\MeV} and energy spread of 19~\si{\MeV}. Electron injection occurred later in the plasma, around 1.8~\si{\mm}, as the laser was focused $\approx1.5$~\si{\mm} upstream of the gas cell entrance, at the start of the density profile. During propagation, the laser initially defocused before refocusing within the plasma, where self-focusing dominated and the ionization injection threshold ($a_0 \approx 2$) was reached, leading to electron trapping in the wake.

\begin{figure*}
    \includegraphics[width=0.45\linewidth]{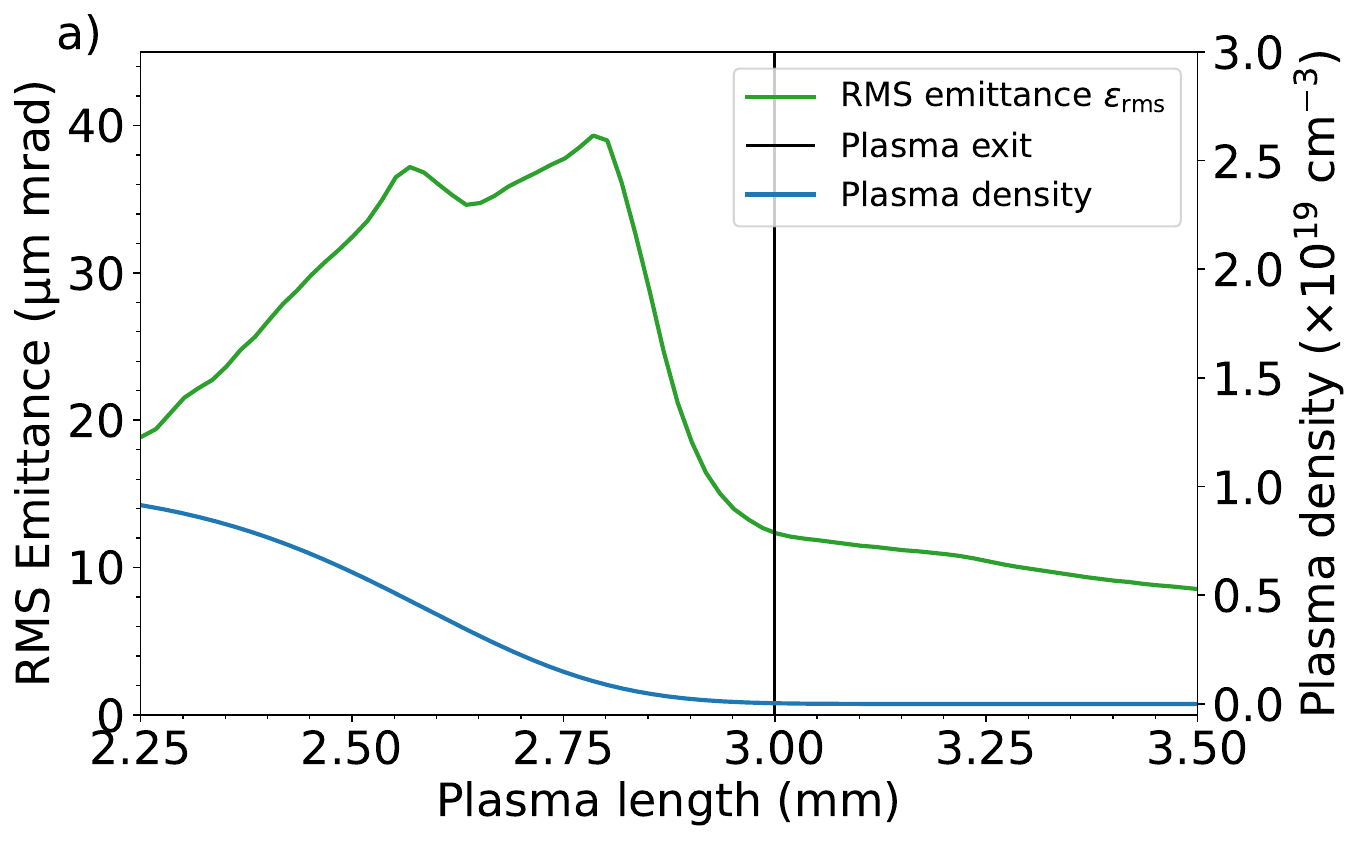}\quad
    \includegraphics[width=0.45\linewidth]{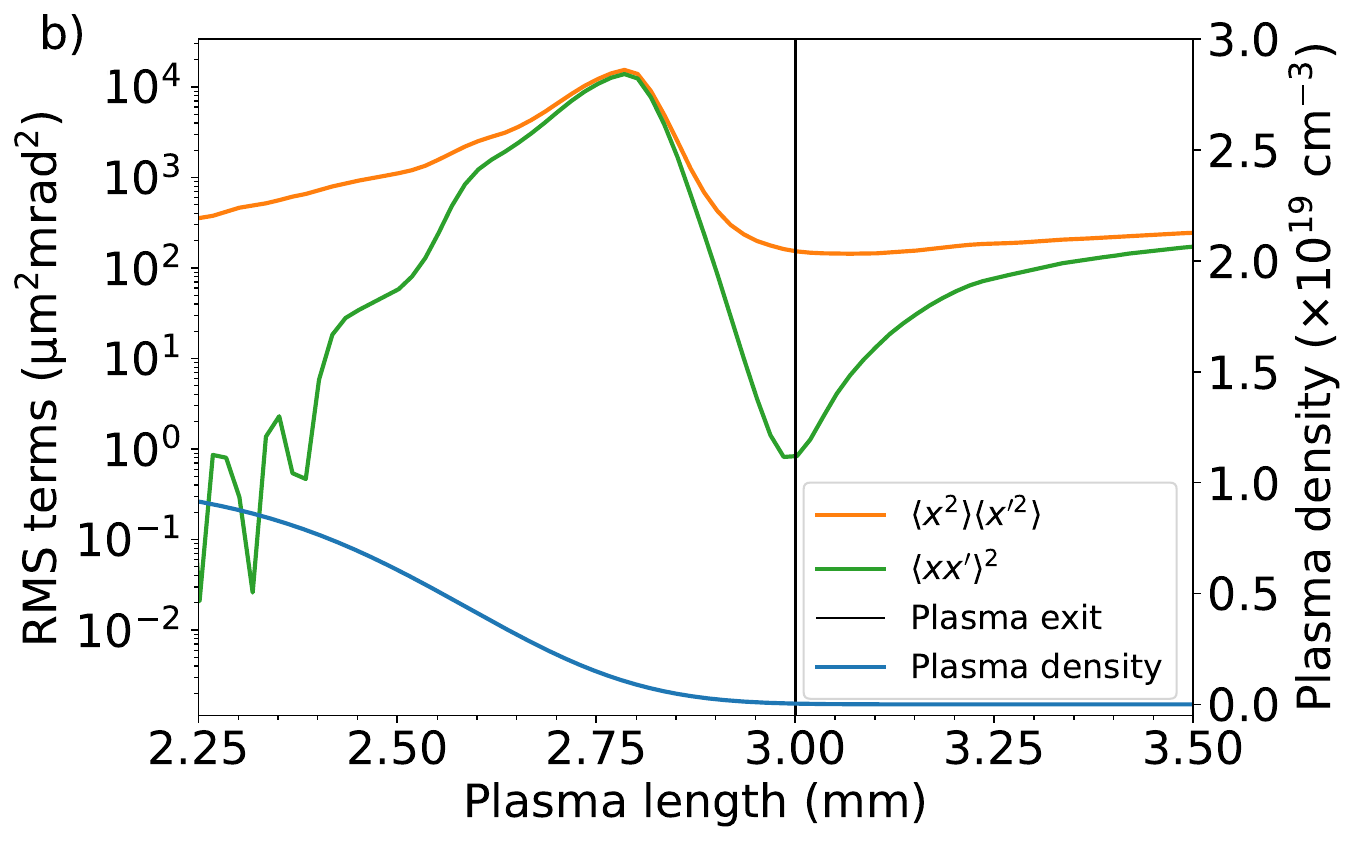}
    \caption{Particle-in-cell simulation result showing the evolution of the (a) RMS emittance and (b) RMS terms of the electron beam at the plasma-vacuum interface which is located around 3~\si{\mm}, where the plasma density goes to zero.}
    \label{fig-appendix:pic-rms-evolution}
\end{figure*}

The simulation provides access to various electron beam parameters as it propagates through the plasma and across the plasma–vacuum boundary, including the evolution of the RMS quantities $\langle x^2 \rangle$, $\langle x^{\prime 2} \rangle$, and the cross-correlation term $\langle x x^{\prime} \rangle^2$ at different positions. Figure~\ref{fig-appendix:pic-rms-evolution} shows the evolution of these parameters extracted from the simulation. At the plasma exit, located at approximately 3~\si{\mm}, where the plasma density approaches zero, both the RMS emittance and beam divergence become constant as the beam propagates into vacuum. Furthermore, at this position, the cross-correlation term $\langle x x^{\prime} \rangle^2$ is about two orders of magnitude smaller than $\langle x^2 \rangle \langle x^{\prime 2} \rangle$, validating the approximation used in Eq.~\eqref{eq. rms emittance definition}, where the cross-correlation term was neglected. This indicates that the beam is effectively at a waist at the plasma exit. The values of the beam parameters corresponding beam and Twiss parameters ($\alpha$, $\beta$, and $\gamma$)~\cite{Reiser.2008} at this plasma-vacuum interface position are summarized in Table~\ref{tab-appendix-plasma-exit}.

According to the simulations, the RMS emittance at the plasma down-ramp exit is about $12.69~\si{\um\mrad}$ and stays roughly constant as the beam travels in vacuum. At the plasma exit, the simulated source size is about $\approx1.9~\si{\um}$, which is consistent with our earlier result of $(1.7\pm0.2)~\si{\um}$ reported in Ref.~\cite{Salgado.2024} for the same setup. The simulated RMS emittance also agrees with our previously reported value of $(13.8\pm2.8)~\si{\um\mrad}$ from a similar setup, though that one was obtained using a different method called laser grating~\cite{Salgado.2024}.

\begin{table}[ht]
\caption{Electron beam parameters at the plasma exit according to our PIC simulations. The first RMS term in the geometric emittance definition from Eq.~\eqref{eq. rms emittance definition}, $\langle x^2 \rangle \langle x^{\prime 2} \rangle$, is about two orders of magnitude larger than the cross-correlation term $\langle x x^{\prime} \rangle^2$, showing that the beam is at a waist at this point.}
\label{tab-appendix-plasma-exit}
\begin{tabular}{lc} \toprule
{Beam Parameters} & {Value at plasma exit} \\ \midrule
$\langle x^2 \rangle \langle x^{\prime 2} \rangle$ & $161.90~\si{\um^2\mrad^2}$ \\ 
$\langle x x^{\prime} \rangle^2$ & $0.82$~\si{\um^2\mrad^2} \\
RMS divergence $\theta_x$ & 6.75~\si{\mrad} \\ 
RMS Beam size $\sigma_x$ & 1.89~\si{\um} \\ 
$\epsilon_{\mathrm{rms},\,x}$ & 12.69~\si{\um\mrad} \\ 
$\alpha {= - \langle x x^{\prime} \rangle / \epsilon_{\mathrm{rms},x} }$ & $-0.0713$ \\ 
$\beta { = \langle x^2\rangle / \epsilon_{\mathrm{rms},x} }$ & $2.802\times10^{-4}$~\si{\m} \\ 
$\gamma { = \langle x'^2\rangle / \epsilon_{\mathrm{rms},x}  }$ & 3586.78~\si{\meter^{-1}} \\ \bottomrule
\end{tabular}
\end{table}

\section{Fitting parameters for the Monte Carlo simulations}
\label{appendix:fitting}
In Fig.~\ref{fig:source scan geant4}, curves of the form $\epsilon_{\rmscalc}^2 = \Omega^2\theta_{x}^2 + \epsilon_{\rmsin}^2$ were fitted to the data points from the emittance evaluation of the Monte Carlo simulations performed for various pepper-pot masks with two different hole pitch distances. In this fitting curve, the divergence parameter $\theta_{x} = 1.85~\si{\mrad}$ is a known quantity determined by the divergence properties of the beam, $\epsilon_{\rmsin}$ is the independent (free) variable, and $\Omega$ is the fitting parameter that depends on the geometry parameters previously discussed.

The fitting parameters $\Omega$, obtained from the curves fitted between the calculated and input emittance values from the GEANT4 simulations for the two pepper-pot masks with pitch distances of 120~\si{\um} and 150~\si{\um} and various hole diameters, are given in Table~\ref{tab:appendix alpha parameter}.

For the curves fitted to the data points of the pepper-pot mask with a 150~\si{\um} pitch in Figs.~\ref{fig:source scan geant4}b) and c), the fitted $\Omega$ parameters obtained using the penumbra method, along with their theoretical values, are also given in Table~\ref{tab:appendix alpha parameter}. These results show good agreement between the $\Omega$ values obtained from the penumbra fitting and the theoretical constant. Fit errors are within 5\% of the fitted value, as given by the covariance matrix resulting from the curve fit.

\begin{table}[ht]
\caption{Fitting parameters for the curves presented in Fig.\ref{fig:source scan geant4} for the pepper-pot masks with pitch distances of 120~\si{\um} and 150~\si{\um}. The fitted values of the $\Omega$ parameter are also given for the mask with a 150~\si{\um} pitch and hole diameters of 20~\si{\um} and 50~\si{\um}, evaluated using the convolution (penumbra) method, along with the corresponding theoretical values. The $\Omega$ parameter obtained from the penumbra method is in good agreement with the theoretical value.}
\label{tab:appendix alpha parameter}
\begin{tabular}{lcc} \toprule
\multirow{2}{*}{Hole diameter $d$ (\si{\um})}  & \multicolumn{2}{c}{$\Omega$ (\si{\um})} \\ \cmidrule(lr){2-3}
  & \textbf{$\Lambda = 120~\si{\um}$} & \textbf{$\Lambda = 150~\si{\um}$} \\ \midrule
20 & 3.17 & 3.45 \\
30 & 3.43 & 3.42 \\
40 & 3.44 & 3.32 \\
50 & 3.89 & 4.28 \\
20 (penumbra) & --- & 0.86 \\
20 (theory) & --- & 0.86 \\
50 (penumbra) & --- & 2.05 \\
50 (theory) & --- & 2.07 \\ \bottomrule
\end{tabular}
\end{table}

\end{document}